\documentclass{CUP-JNL-DTM}%

\usepackage{graphicx}
\usepackage{multicol,multirow}
\usepackage{amsmath,amssymb,amsfonts}
\usepackage{mathrsfs}
\usepackage{amsthm}
\usepackage{rotating}
\usepackage{appendix}
\usepackage{anyfontsize}
\usepackage{ifpdf}
\usepackage[T1]{fontenc}
\usepackage{newtxtext}
\usepackage{newtxmath}
\usepackage{textcomp}
\usepackage{xcolor}
\usepackage{tcolorbox}
\usepackage{framed}
\usepackage{lipsum}
\usepackage[colorlinks,allcolors=blue]{hyperref}
\usepackage[utf8]{inputenc}

\usepackage[
    backend=biber,
    style=authoryear,
    hyperref=true,
    url=false,
    isbn=false,
    backref=false,
    maxcitenames=2,
    maxbibnames=100,
    block=none
]{biblatex}

\DeclareNameFormat{family-given-bold}{\mkbibbold{%
  \ifgiveninits
    {\usebibmacro{name:family-given}
      {\namepartfamily}
      {\namepartgiveni}
      {\namepartprefix}
      {\namepartsuffix}}
    {\usebibmacro{name:family-given}
      {\namepartfamily}
      {\namepartgiveni}
      {\namepartprefix}
      {\namepartsuffix}}%
  \usebibmacro{name:andothers}}}
\DeclareNameAlias{sortname}{family-given-bold}

\renewbibmacro{in:}{}

\theoremstyle{definition}

\numberwithin{equation}{section}

\jname{Data \& Policy}
\articletype{RESEARCH ARTICLE}
\jyear{2023}

\begin{document}

\begin{Frontmatter}

\title[Article Title]{Honest Computing: Achieving demonstrable data lineage and provenance for driving data and process-sensitive policies}

\author*[1]{Florian Guitton \authororcid{0000-0002-1617-6620}*}
\author[1,2]{Axel Oehmichen \authororcid{0000-0003-0224-5709}}
\author[2]{Étienne Bossé}
\author[3]{Yike Guo}

\authormark{Florian Guitton \textit{et al}.}

\address[1]{\orgdiv{Data Science Institute}, \orgname{Imperial College London}, \orgaddress{\city{London}, \postcode{SW7 2AZ}, \state{Greater London}, \country{United Kingdom}}}

\address[2]{\orgname{Secretarium Ltd}, \orgaddress{\city{London}, \postcode{EC4M 9DN}, \state{Greater London},  \country{United Kingdom}}}

\address[3]{\orgname{Hong Kong University of Science and Technology}, \orgaddress{\country{Hong Kong}, \state{Clear Water Bay}}}

\address*[*]{\text{Corresponding author.\email{f.guitton@imperial.ac.uk}}}

\authormark{Florian L. P. Guitton, Axel F. Oehmichen, \textit{et al}.}


\abstract{
Data is the foundation of any scientific, industrial or commercial process. Its journey typically flows from collection to transport, storage, management and processing. While best practices and regulations guide data management and protection, recent events have underscored its vulnerability. Academic research and commercial data handling have been marred by scandals, revealing the brittleness of data management. Data, despite its importance, is susceptible to undue disclosures, leaks, losses, manipulation, or fabrication. These incidents often occur without visibility or accountability, necessitating a systematic structure for safe, honest, and auditable data management.

In this paper, we introduce the concept of Honest Computing as the practice and approach that emphasizes transparency, integrity, and ethical behaviour within the realm of computing and technology. It ensures that computer systems and software operate honestly and reliably without hidden agendas, biases, or unethical practices. It enables privacy and confidentiality of data and code by design and by default. We also introduce a reference framework to achieve demonstrable data lineage and provenance, contrasting it with Secure Computing, a related but differently orientated form of computing. At its core, Honest Computing leverages Trustless Computing, Confidential Computing, Distributed Computing, Cryptography and AAA security concepts.

Honest Computing opens new ways of creating technology-based processes and workflows which permit the migration of regulatory frameworks for data protection from principle-based approaches to rule-based ones. Addressing use cases in many fields, from AI model protection and ethical layering to digital currency formation for finance and banking, trading, and healthcare, this foundational layer approach can help define new standards for appropriate data custody and processing.
}

\end{Frontmatter}

\begin{tcolorbox}[boxrule=0pt,colback=gray!20]

\section*{Policy Significance Statement}{
Regulatory frameworks have the vocation to create auditable, validatable and enforceable rules in any given process with equal expectations to mandate privacy, security, and fairness. This has been a significant challenge in sensitive data processing and automated decision-making systems.

Large regulatory bodies are limited in their ability to prescribe demonstrable and systematic protocols for handling and managing these data and processes while ensuring transparency at scale. Indeed, they prefer to define high-level principles and shy away from defining ownership structures. This is largely due to a lack of technical readiness and feasibility at the time of writing.

Honest Computing, rooted in trustless computing and confidential computing paradigms, promises to remedy complex regulatory gaps by providing a technological basis for action, thereby empowering policymakers to establish more transparent and effective guidelines to ensure compliance, transparency, accountability, and ethical conduct across all verticals.

}
\end{tcolorbox}


\section{Introduction}\label{sec:intro}

In the digital era, policymakers grapple with an array of complex challenges when it comes to effectively regulating data management. The exponential growth in data volume, coupled with the explosion of data modalities, presents a formidable hurdle for crafting comprehensive and adaptable regulatory frameworks. The primary goal of regulatory frameworks in any domain, particularly in data protection and privacy, is to establish a structured and protective environment that safeguards individuals, organisations, and societal interests.  Nevertheless, these frameworks must equally foster innovation and economic growth and strike the correct balance between responsible handling and scope of data utilisation.

Naturally, policymakers must reckon with the usual tensions surrounding the fear of new technologies, particularly as the complexity threshold extends beyond general knowledge and understanding. This is particularly true of use cases that call upon concerns of mass surveillance, something which was explicitly evident during the COVID-19 pandemic (\cite{eck_state_2020}), which saw governments implement contact-tracing apparatuses but also in the context of telecom data analysis (\cite{oehmichen_opal_2019}).

In this article, we motivate and present a specification of Honest Computing as a fully-fledged off-the-shelf solution for ethical data privacy. It leverages a modern, unique assembly of technologies which have recently reached technological readiness in the industry and are available on the market today. We also draw attention to the contrast between Honest Computing and Secure Computing, emphasising the difference in focus between these two adjacent concepts.

\subsection{Motivation}

While data protection regulations aim to fortify individuals' privacy rights and enhance data security, they often introduce complexities and trade-offs impacting various facets of the digital landscape. Stricter regulations may impose financial burdens, particularly on smaller businesses or startups, necessitating investments in compliance measures and potentially hindering innovation. Moreover, stringent rules impede data sharing and interoperability among organisations, limiting the potential for collaborative research or inhibiting the development of new services. Conversely, lax regulations compromise individuals' privacy and expose them to data breaches or exploitation by entities seeking to profit from personal information. Finding the equilibrium between protecting privacy and enabling innovation, between stringent regulations and fostering growth, stands as a fundamental trade-off in crafting effective data protection measures.

To make matters more complex, the continuous digitalisation of everything represents a sweeping transformation across industries, societies, and daily life, driven by integrating digital technologies into every aspect of our world. This further commands the pervasive use of digital tools, data-driven systems, and interconnected networks that redefine how we communicate, work, interact, learn and discover. From smart devices and IoT (Internet of Things) sensors to automated processes and AI-driven decision-making, the digitalisation of everything reshapes traditional norms, jumps across geographical boundaries and adds a scalability dimension to policymakers' work. On the matter of AI, while it applies to any model generated from data, model defence and intellectual property protection are also topics of great concern for policymakers (\cite{picht_ai_2023}).

It is also helpful to consider the different classes of regulations that policymakers leverage to crystallise these strict requirements. Rule-based and principle-based regulations represent two distinct approaches to establishing regulatory frameworks, each with its characteristics and applications. Rule-based regulations are specific, prescriptive, and detailed, laying out explicit guidelines, procedures, and requirements for compliance. These regulations leave little room for interpretation, offering clear-cut directives and leaving minimal discretion to those governed by the rules. They are well-suited for addressing straightforward situations; providing clarity and consistency in enforcement. On the other hand, principle-based regulations are broader and more flexible, emphasising overarching principles, values, and goals rather than detailed instructions. These principles set general standards and objectives, offering a framework within which stakeholders exercise judgment and discretion to tailor their actions to achieve compliance. Principle-based regulations foster adaptability and innovation, allowing for more dynamic responses to evolving circumstances. While they offer greater flexibility, they might also introduce ambiguity and require more interpretation, potentially leading to inconsistencies in enforcement. Balancing the strengths and weaknesses of both approaches is crucial in crafting effective regulatory frameworks that achieve compliance, promote fairness, and cater to the dynamic needs of diverse industries and societal contexts. With this context at hand, we understand how regulations like the European Union's General Data Protection Regulation (GDPR), through lack of technological readiness, have preferred to approach the issue of sensitive data handling with a more principle-driven approach, leading to its current gaps.

The complexity of validation and enforcement hinders ideal data handling features such as the processes' transparency and their exhaustiveness of description. This is a two-part problem again. On the one hand, understanding the data and keeping it safe from unauthorised reads, potential alteration, accidental deletion or error is not enough. On the other hand, we understand that keeping track of what has happened to the data and the exact processes it has gone through is an unanswered problem. Driving the accountability of all actors in the data journey is a complicated task and today almost entirely relies on manual processes prone to human error, black-box systems with unsatisfactory protections subject to manipulation by administrative or otherwise privileged personnel, and altogether out of reach of responsible authorities mandated to perform audits only after a breach has occurred. This is compounded by valid constraints on audit by authorities due to other regulations on preserving industrial secrets and adherence to other standards defining strict internal perimeters for private companies to perform contractual obligations, particularly if these audits ought to be routine rather than compelled by a judiciary act. Many cases exist in the records of companies willingly disposing of or concealing incriminating information (\cite{the_associated_press_ex-uber_2023,townsend_facebook-cambridge_2022,bitdefender_bitdefender_2023}).

The constantly evolving technological landscape outpaces the speed at which regulations can be formulated and implemented with appropriate context, creating a persistent struggle to keep pace with emerging data-related risks and privacy threats, as such data protection regulations feature a series of trade-offs that policymakers and stakeholders must carefully navigate. Paradoxically, until very recently, there were no practical technological solutions to these problems amplified by the progress of digitalisation. The door is ajar, and since 2015, an array of new technologies have been developed, which can be assembled into systems fit to empower the most prescriptive regulatory frameworks.

\subsection{Background}

To provide appropriate solutions, however, it is fundamental to understand the key features of the challenges policymakers face when drafting regulatory frameworks and the critical requirements that new technological systems must exhibit to solve them. When approaching the drafting of regulation, policymakers must understand the data pipeline, from collection to transport, storage, management, processing, protection and disposal.

Data pipelines involve a complex chain of interconnected activities that starts with a data source and ends in a data sink. Data pipelines are important for data-driven organisations since a data pipeline can process data in multiple formats from distributed data sources with minimal human intervention, accelerate data life cycle activities, and enhance productivity in data-driven applications. However, in practice, raw data are rarely ready to be consumed and must be transformed by a succession of operations usually referred to as data pipelines. There are many reasons why a data source cannot be used directly. For instance, if too many descriptive variables exist, some feature selection or dimensionality reduction algorithms must be applied. All those operations introduce bias, and their presence or absence in a data pipeline may be subject to discussion. The data pipeline depends both on the data source and the algorithm, such that no universal pipeline can work for every data source and algorithm. As such, a technological answer to the policymaking challenge must retain a high degree of flexibility. We can, however, isolate two distinct primary principles to consider for any well-controlled data pipeline: data provenance and data lineage. Data provenance and lineage serve as fundamental principles in ensuring better data protection by offering critical insights into the origin, movement, and transformation of data throughout its lifecycle.

\subsubsection{Data provenance: Origin and authenticity of the data}
Data provenance — sometimes called ``pedigree'' — describes the origins of a piece of data and the process by which it makes its way into a database system. With the proliferation of specialist databases and curated catalogues, the issue of data provenance - where a piece of data came from and the process, sometimes the whole ETL (Extract, Transform, Load) pipeline, by which it arrived in the database - is becoming increasingly important, especially in scientific databases where understanding provenance is crucial to the accuracy and currency of data (\cite{buneman_why_2001}). Data provenance involves tracing the history of data elements, detailing their origins and any changes they undergo. This information is vital for verifying data authenticity, reliability, and integrity, thereby enhancing trustworthiness and reducing manipulation or tampering risks. Understanding data provenance enables better decision-making regarding data quality and credibility (\cite{buneman_data_2018}).

\subsubsection{Data Lineage: The information journey} 

Similarly, data lineage encompasses the complete journey of data, depicting how it moves, changes, and transforms as it travels through various systems, processes, and organizations. This lineage helps in understanding dependencies, identifying potential vulnerabilities, and assessing the impact of changes or modifications on the overall data integrity and security.

In its most general form, lineage describes where data came from, how it was derived, and how it was updated over time. Information management systems today exploit lineage in tasks ranging from data verification in curated databases to confidence computation in probabilistic databases. Lineage can be helpful in a variety of settings. For example, molecular biology databases mostly store copied data and can use lineage to verify the copied data by tracking the original sources. Data warehouses can use the lineage of anomalous view data to identify faulty base data, and probabilistic databases can exploit lineage for confidence computation. In addition to challenges related to space and time efficiency, it can be difficult to define lineage in domains that allow arbitrary transformations (\cite{ikeda_data_2009}).

By implementing robust data provenance and lineage practices, organizations can effectively track data movements, detect anomalies, and ensure adherence to regulatory requirements such as GDPR. These principles facilitate accountability and transparency, empowering organizations to demonstrate compliance, mitigate risks related to data breaches or unauthorized access, and promptly address privacy concerns. Data provenance and lineage must work closely together to provide an ``honest'' origin of the data.

Furthermore, data provenance and lineage are pivotal in enabling effective incident response and forensic analysis during security breaches or data incidents. They provide crucial information for investigations, helping organizations identify the source of breaches and take appropriate remedial actions. In recent history, the veracity of data in research has been put into question (\cite{dunleavy_is_2023}) due to the same lack of policymakers' experience.

\subsubsection{Technological Readiness Levels}
Technology Readiness Levels (TRLs) are a classification used to assess the maturity and readiness of a particular technology or innovation. There exist nine levels, each representing a stage in the development and validation process of a technology. Developed by NASA in the 1970s to evaluate space technologies, TRLs have been widely adopted by various industries, including aerospace, defence, healthcare, and more, as a standardized method to gauge the readiness of technologies for implementation or commercialization (\cite{mankins_technology_1995}). Assessing technologies using TRLs helps stakeholders, including researchers, investors, policymakers, and industries, understand the current state of technology development, estimate risks, and make informed decisions regarding investment, further development, or deployment. It also assists in communicating the maturity of technologies across different stages of development, fostering collaboration and innovation between researchers, industries and regulators (\cite{olechowski_technology_2015}). In this article, we postulate that to be an effective tool for policymakers, our specification for a reference architecture of Honest Computing must, at the very least, be based exclusively on technologies that have achieved a TRL 5 or above, meaning technologies which have already been successfully tested outside of a lab in real-world conditions.

\subsubsection{Digital Evidences}

Digital evidence refers to electronic data or information collected, preserved, and analyzed to support investigations or legal proceedings. It is a particular element in the policy landscape, as understanding and validating its validity has an intrinsic legal value. Digital evidence is crucial in guaranteeing chains of custody, investigating security breaches or complying with legal proceedings. However, despite significant advancements in cybersecurity in the age of AI, authenticating digital evidence remains a complex and challenging task (\cite{santamaria_smart_2023}). Existing work on digital evidence provides another base of knowledge and constraints on which to build the requirement for Honest Computing. In other words, an Honest Computing system must be able to provide guaranteed, validated digital evidence.

\subsection{Requirements}

To provide an incentive for use, the design of a prescriptive Honest Computing system must possess certain non-functional qualities aiding its adoption. We first postulate that such a system must be Turing-complete. Turing completeness is a concept used in computer science to describe a system's capability to perform any computation that a Turing machine can, given enough time and memory. A Turing-complete system or programming language can simulate a Turing machine, meaning it can execute any algorithm or compute any computable function. Such a system must also be fast; with the ever-growing size of datasets and computation complexity, particularly with AI workloads, an ideal solution would be able to reach reasonable performance indicators.

Finally, Honest Computing achieves accountability through three critical pillars: transparency, integrity, and verifiability. It requires demonstrating transparency via verifiable lineage while providing security and confidentiality.

\subsubsection{Transparency}
Transparency in software systems refers to the accessibility and visibility of information regarding the functioning, processes, algorithms, and data handling practices within software applications. It encompasses the ability for stakeholders, including users, developers, and regulatory bodies, to understand how the software operates, processes data, and impacts it has on users and society at large (\cite{leite_software_2010}). Transparency in software systems holds significant implications for policy challenges, especially in the realms of privacy and accountability (\cite{ferraiuolo_policy_2022}).

When software operates in opaque or proprietary ways, accountability becomes challenging. Transparent systems enable accountability by allowing stakeholders to trace the decision-making processes, identify potential biases, and understand the implications of software-driven actions. Establishing policies that promote transparency in algorithmic decision-making and AI systems becomes imperative to hold developers and organisations accountable for the outcomes of their software (\cite{portugal_is_2017}).

Ethical considerations in technology also intersect with transparency in software systems. Policies addressing the ethical use of software often focus on transparency to ensure fairness, prevent algorithmic biases, and uphold societal values. Transparency allows for scrutiny and evaluation of software systems to detect and address potential ethical issues, such as discrimination or manipulation, fostering ethical accountability among technology providers (\cite{maggiolino_eu_2019}). In fact, recent work in the field of AI model classification reinforces the need for a better understanding of fine-grained transparency throughout, proposing a nomenclature for its evaluation (\cite{bommasani_foundation_2023}).

However, achieving transparency in software systems presents its challenges. Proprietary software, trade secrets, or complex algorithms might hinder full transparency. Balancing the need for transparency while protecting intellectual property rights and commercial interests poses a significant policy dilemma. Striking a balance between fostering innovation and ensuring transparency and accountability requires nuanced policy frameworks that encourage transparency without stifling innovation or hindering competitiveness.

In essence, transparency in software systems is essential for empowering users, fostering accountability, and addressing ethical concerns in technology. Policymakers must navigate the complexities of incentivising transparency while preserving innovation to create frameworks that promote responsible and transparent use of software systems.

\subsubsection{Verifiability}
Verifiability confirms and validates the correctness and reliability of computing processes and outcomes. It involves providing mechanisms and tools to independently verify the accuracy and integrity of computations, data, and results. This pillar is essential for building policies that can render computing systems accountable.

Cryptographic attestations are cryptographic proofs or assertions particularly suited to provide verifiable evidence or confirmation about the integrity, authenticity, or specific attributes of data, software, or hardware components. They play a crucial role in ensuring trust, security, and verifiability in various digital systems, especially in distributed environments such as blockchain networks, cloud computing, or IoT (Internet of Things) ecosystems.

These attestations are created using cryptographic techniques that generate digital signatures or hashes linked to specific pieces of information. They are used to verify the identity of entities, confirm the integrity of data or software, or attest to the correctness of certain operations or configurations. Cryptographic attestations in aspects relevant to designing data systems for policymaking can be categorised into different types:

    \begin{itemize}
        \item \textbf{Code Signing:} Cryptographic signatures, used in software distribution, authenticate the origin and integrity of software packages. Code signing certificates are used to sign software binaries, ensuring they have not been tampered with and come from a trusted source (\cite{de_carne_de_carnavalet_challenges_2014}). This is a useful tool in the data supply chain to provide a sane base for data provenance. Over the past few years, actors such as The Linux Foundation\textsuperscript{\textregistered} and their Sigstore project have provided developers with streamlined and free infrastructure to leverage such techniques.

        \item \textbf{Blockchain-based Attestations:} In blockchain networks, attestations are used to verify the validity of transactions or data. Cryptographic proofs are generated and stored on the blockchain, providing a transparent and immutable record of actions or events (\cite{aydar_towards_2020}).

        \item \textbf{Remote Attestation:} This type of attestation is commonly used in trusted computing environments to verify the integrity of a remote device or platform (\cite{dushku_prove_2023}). A remote entity can generate cryptographic evidence attesting to the security and integrity of its software or hardware components (\cite{lie_specifying_2003}). A trusted party then verifies this evidence to ensure that the remote device is operating securely and has not been compromised. Trusted Platform Modules (TPM) are a type of device using a hardware-based security module to store cryptographic keys and generate a unique attestation identity key (AIK) for a device. Using the AIK, a device can prove its identity and integrity to a remote verifier. Trusted Execution Environments such as Intel\textsuperscript{\textregistered} SGX have the capacity to provide such features in an integrated fashion as well (\cite{shepherd_secure_2016}).
    \end{itemize}

Cryptographic attestations bolster trust by allowing stakeholders to verify the legitimacy, integrity, and authenticity of data, software, or hardware components without relying solely on trust in a centralised authority. These attestations leverage cryptographic techniques to provide verifiable proof of various properties, ensuring that digital interactions and systems can be trusted and validated in decentralised and distributed environments.

\subsubsection{Integrity}

Also referred to as tamperproofness, there are two main definitions of integrity in the context of data management, akin to two sister concepts of tamperproofing.

The first definition refers to data integrity, which is the accuracy and consistency of data over its entire lifecycle. Ensuring data integrity requires implementing appropriate data validation processes, maintaining proper documentation, and implementing security controls to prevent unauthorized access, modification, or deletion of data. This is particularly important for sensitive data, such as financial or personal information, where the accuracy and completeness of the data are critical. Data integrity mechanisms are tamper-evident in nature, and appropriate validation mechanisms will alert stakeholders.

For instance, a stakeholder could choose to use a Merkle Patricia Trie (MPT) to provide data integrity. MPTs are a type of data structure, an enhancement of the original Merkle Tree data structure, used in blockchain technology to store and retrieve data efficiently in a cryptographically secure manner. They are, for example, employed in the Ethereum blockchain, providing an efficient way to store and retrieve the state of an Ethereum account and its associated data (\cite{mardiansyah_multi-state_2023}).

The second definition of integrity in data management refers to system integrity, which is the overall reliability and consistency of a system. System integrity encompasses a range of factors, including hardware and software reliability, network availability, and data backup and recovery measures. Ensuring system integrity requires implementing robust security measures, such as firewalls, intrusion detection systems, and access controls, to prevent unauthorized access and protect against system failures or data loss. In the context of Honest Computing, we approach goals of system integrity as tamper-resistance.

As it pertains to computing and technology, tamper-proofing usually refers to the tamper-evident nature of a process, data or device. A typical example would be Hardware Security Modules (HSMs); they are dedicated hardware devices designed to protect and manage cryptographic keys and operations and built with physical security mechanisms, such as strong enclosures and sensors, to detect and respond to tampering attempts.

Integrity and tamper-evidentness are essential aspects of any effective data management to bring about solid bases for regulatory framework development. By incorporating robust and secure technologies and techniques, policymakers can ensure the accuracy, reliability, and trustworthiness of data while also mandating the prevention of unauthorized or malicious alteration of data that could lead to potential harm or misuse. Additionally, it is worth mentioning that a technological solution to data integrity must follow standard best practices and ensure resistance to hardware failure other than security-related. An honest Computing system must be reliable and ensure appropriate data replication.

\subsubsection{Deterministic Execution}
Deterministic execution refers to a computing process or system's ability to produce the same output or result when given the same input and operating under the same conditions every time it runs. In other words, a deterministic system will always yield identical outputs for a specific set of inputs and environment, regardless of the number of times it executes. There is a consensus in the research community that non-determinism makes the development of parallel and concurrent software substantially more difficult (\cite{bergan_deterministic_2011}), and this difficulty carries over into the attestation of such software. Studies in the computational reproducibility of research work also encourage loudly that such practice must be the standard (\cite{cruwell_whats_2023}).

Key characteristics and significance of deterministic execution:

\begin{itemize}
    \item \textbf{Reproducibility:} Deterministic systems ensure the reproducibility of results, which is crucial in scientific research, simulations, and testing environments. It allows researchers to validate findings and verify experimental results.

    \item \textbf{Consistency:} Deterministic execution guarantees consistency across different executions, even in distributed systems. This consistency is vital in ensuring that all nodes in a distributed network reach the same state given the same inputs.

    \item \textbf{Debugging and Testing:} Deterministic systems simplify debugging and testing processes. Identical outcomes for a given input facilitate easier identification and resolution of issues.

    \item \textbf{Predictability:} Predictable behaviour in software and systems aids in understanding and anticipating their operation, which is essential in critical applications like finance, healthcare, and aerospace.

    \item \textbf{Cryptographic Operations:} Deterministic execution is crucial in cryptographic operations. Cryptographic algorithms must produce consistent outputs to ensure security and validate transactions across distributed networks like blockchains.
\end{itemize}

In software development and system design, efforts must be made to achieve deterministic behaviour whenever possible, especially in critical systems where predictability, reliability, and consistency are paramount. Techniques like using deterministic algorithms, avoiding randomness in critical paths and ensuring synchronisation mechanisms contribute to achieving deterministic execution in software and systems. For applications that require a degree of randomness, for example, it is imperative to implement distributed random generation algorithms (\cite{lalev_distributed_2018}).

\section{Threat Model for Honest Computing}\label{sec:proposedarchitecture}

In this section, we present a formal threat model for Honest Computing, outlining common attack scenarios, identifying vulnerabilities, and proposing mitigation strategies to enhance computing systems' honesty posture. Threat modelling is a common practice in software engineering development; however, we see that up to 73.68\% of the assumptions made in these modelling exercises are meant to exclude threats from consideration (\cite{vanlanduyt_descriptive_2022}). The study also found that less than 10\% of assumptions referred to the attackers' abilities, potentially leaving substantial gaps in threat mitigation and compromising the soundness of security architecture design. A correctly executed Honest Computing implementation could be a validatable foundation for safe threat exclusion.

\subsection{Framing}

In this article, we are not concerned with attack vectors susceptible to exposing end-user data but rather attack vectors susceptible to compromising a system's privacy, honesty or its ability to demonstrate privacy and honesty. One of the intrinsic goals of Honest Computing is to offer a privacy-enabling and trustless/zero-trust architecture.

\subsubsection{Honesty vs Security}

It is important to understand the differences between ``Honest Computing'' and ``Secure Computing'' in order to correctly assess a viable threat model for Honest Computing. Indeed, secure computing is not necessarily intrinsically honest. While secure computing aims to protect computing systems from unauthorised access and disruption, it does not inherently ensure honesty or transparency in how those systems operate or how data is used.

On the one hand, secure computing primarily focuses on technical measures such as encryption, access controls, and intrusion detection to prevent security breaches and maintain the integrity of computing systems. However, it does not guarantee that the system's behaviour is always honest or that users have complete visibility into how their data is processed, or decisions are made.

On the other hand, Honest Computing emphasizes transparency, accountability, traceability and ethical use of technology. It involves practices such as providing clear explanations of data processing methods, disclosing potential biases in algorithms, and ensuring that users have a comprehensive understanding of how their data is handled.
While secure computing and Honest Computing share common goals of building trust and confidence in computing systems, they address different aspects of trustworthiness. Secure computing focuses on technical safeguards against security threats. In contrast, Honest Computing addresses broader considerations of transparency, fairness, and ethical use of technology.

\subsubsection{Responsability boundary}

It is equally essential to distinguish between the effectiveness of an Honest Computing system in guaranteeing the safety of the digital evidence it produces and the adequate safety of a user's data. We assume that the applications running on top of our Honest Computing system are ideally programmed and do not contain code that would otherwise handle data in an unsecured fashion (e.g., an API that would simply send data to a remote user outside of its containment unit). As such, we consider the code of software running on the Honest Computing system and accessed by the client to be out of scope. We can, however, provide privacy guardrails to ensure the confidentiality of processing.

\subsubsection{Root of trust}

A root of trust is a foundational component in computer security that establishes a secure starting point for a system. It is a trusted entity or mechanism that serves as the basis of trust for other components or processes within a computing environment. The concept is crucial for ensuring data and processes' integrity, confidentiality, and authenticity in various computing systems, including computers, servers, mobile devices, and embedded systems. There exist multiple levels of root of trust: Hardware-based Root of Trust, Firmware-based Root of Trust, Software-based Root of Trust, Network-based Root of Trust and Human-based Root of Trust. ``Trust'' and ``Honesty'' are closely related concepts but are not synonymous. Trust is dynamic and belief-based rather than deterministic and evidence-based. There are numerous examples of ``Abuse of positions of trust'' where people learnt a posteriori of breaches, cases such as the Barclays Dark Pool scandal (\cite{inman_barclays_2014}) or the NHS Deepmind scandal of 2021 (\cite{bbc_deepmind_2021}). In the context of Honest Computing, we defined that any Root of Trust at the level of the software or above is unsuitable because it cannot be used to rigorously demonstrate digital evidence remotely.

\subsubsection{Technological stack}

In this article, we take the example of an Honest Computing service, hosted on a third-party cloud system, being accessed from a client device via a standard HTTP web call such as a RESTful API, but the reflection is applicable more generally.

\subsection{Methodology}

While recent frameworks such as Yacraf (\cite{ekstedt_yet_2023}) for risk-based threat modelling have been developed, these risk-based models require understanding in great detail the business problem a particular software is attempting to solve and the cost associated. This approach is unsuitable for analysing Honest Computing as its features adopt those of cloud underlays. According to the systematic literature review on threat modelling performed by Xiong et al. (\cite{xiong_threat_2019}), the field of threat modelling in cyber security solutions lacks common ground and mainly comprises extensively manual analysis methods to identify threats. Additionally, we find that describing complex distributed systems providing effective Data Flow Diagrams (DFD) is challenging due to the complementary deployment model nature of these processes (\cite{vanlanduyt_descriptive_2022}). The survey by Hong et al. (\cite{hong_systematic_2019}), which adopts a combinatorial approach between the STRIDE threat categories and the OWASP attack families, provides an excellent example of modern threat modelling.

We have chosen to follow a top-down approach to threat discovery and definition based on STRIDE definitions, starting from the client down to the metal. We assume that at every level, third parties are malicious or untrustworthy.

\subsection{Threat chain and mitigation strategies}

\begin{itemize}

    \item \textbf{S1 - Compromised client processing:} An attacker has compromised the client device and is injecting false responses to the ingress. \textit{Mitigation: Implement a system based on remotely attestable digital artefacts that can be independently checked against a discreet system.}
    
    \item \textbf{I1 - Man-in-the-middle:} A proxy service or boundary server with legitimate domain ownership terminates the TLS connection from an upstream server and re-encrypts in egress. As a consequence, the proxy can eavesdrop or alter the content of the communication (\cite{bhargavan_formal_2018}). \textit{Mitigation: A double TLS encryption is initiated to protect the data flow from the TLS terminations. The client no longer relies on HTTP host validation.}
    
    \item \textbf{I3 - Software leak:} The software developer manipulates your data in memory and extracts insight without your knowledge. (\cite{castes_attestable_2022}) \textit{Mitigation: Ensure that the code is remotely attestable by being run in a TEE.}
    
    \item \textbf{I4 - Runtime inspection:} The platform runtime executing within a TEE enclave implements a backdoor which allows it to spy on the communication between the client and the software post-TLS termination. \textit{Mitigation: Ensure that the runtime manager and the software run in separate enclaves while implementing a two-layer encryption system within the double TLS connection with the client.}
    
    \item \textbf{E1 - Theft of the TEE key:} An attacker performs a lab attack to extract the TEE key (e.g. through using an electron microscope). \textit{Mitigation: The TEE key should not be used directly as the sole means to secure data and code in memory, the runtime must rely on a distributed rotating key to store data in memory or down to disk. Multiple TEEs could exploit distributed ledger technology to achieve this. }
    
    \item \textbf{E2 - TEE vendor attack:} The manufacturer of the TEE performs a remote update to the processor microcode to be able to issue erroneous quotes. \textit{Mitigation: Adopting a multi-vendor approach allows to perform horizontal validation that computation was not tampered with. This eliminates the need for a singular Firmware-based Root of Trust by extending the responsibility to a distributed Hardware-based root of Trust. }

    \item \textbf{T2 - Destruction of the TEE:} An attacker attempts to physically damage the TEE in an attempt to remove a trace of their actions. \textit{Mitigation: Adopt multi-cloud MaaS approach ensures that simultaneous destruction of all the TEEs of a cluster is difficult to achieve.}
    
    \item \textbf{T3 - Time shift attack on the TEE:} An attacker attempts to artificially alter the execution speed of the TEE through external factors (e.g. increase the temperature). This is done to artificially force a system to execute a transaction ahead of time. \textit{Mitigation: Implement a time-sensitive consensus protocol for deterministic execution able to detect drift in participants' clock speed to exclude them from the quorum.}
    
    \item \textbf{D1 - TEE side channel attack:} An attacker attempts to corrupt the communication between the secure and normal worlds. This can affect cross-TEE communication and impact the ability of the cluster to validate transactions. (\cite{van_schaik_sok_2022}) \textit{Mitigation: Implement an error-sensitive consensus protocol for deterministic execution able to detect abnormal communication with participants to exclude them from the quorum.}
    
    \item \textbf{S2 - TEE cluster intrusion:} An attacker attempts to inject a compromised TEE in the consensus cluster to steal the shared private key. \textit{Mitigation: Implement a mechanism of positive identification of trusted parties, where the TEEs remote attest each other in an Indentificate Friend or Foe (IFF) fashion (\cite{chen_mage_2022})}
    
\end{itemize}

 Our threat model lays a foundation for understanding the unique challenges inherent in ensuring honesty and privacy within computing systems by delineating responsibility boundaries and elucidating the Root of Trust concepts. The identification of potential threats and corresponding mitigation strategies provides a road to specification and practical implementation of requirements. Looking ahead, continual refinement of the threat model, particularly with the ability to protect further edge and client devices with their own TEEs, will be crucial in adapting to evolving technological landscapes and emerging threats.

\section{Honest Computing reference specifications}\label{sec:honestcomputing}

We propose a first definition of Honest Computing as the practice and approach that emphasises transparency, integrity, and ethical behaviour within the realm of computing and technology. It ensures that computers and software operate honestly, transparently and reliably; and that they enable privacy and confidentiality of data and code by design and by default.

The notion of honesty in computing has been thoroughly studied in Secure Multi-Party Computing (SMPC) research. In his 1998 manuscript, Goldreich defines the outline of semi-honest computing agents and the challenge they represent in that they are assumed to behave according to a defined protocol yet extract and store historical information as they go (\cite{goldreich_secure_1999}).

Honest Computing ought to solve policy challenges by employing technologies that concern themselves with the safety of the data and the safety of the code and process applied to the data. This is why we propose that an Honest Computing system must be built on top of a distributed ledger capable of storing both code and data as appropriate, itself built on top of a confidential computing network of agents. This unique combination permits the provenance gathering of the entire chain of data and workflow, permitting the demonstration of accountability fit for audit and policy enforcement. Of course, in some cases, a stakeholder may choose to store data outside of the ledger to avoid the issue of inappropriate retention as well as integrate non-honest computing services via API, but would then forego the guarantees offered by such a system.

In consideration of the threat model detailed in the previous section, we detail the foundational bricks of Honest Computing.

\subsection{Confidential computing}

Our reference specification proposes to build Honest Computing on top of trusted execution environments (TEEs), such as Intel SGX (\cite{costan_intel_2016}), ARMv9 CAA (\cite{fox_verification_2023}) or RISC-V Keystone (\cite{lee_keystone_2020}). These are hardware-based security solutions that create Turing-complete-capable secure enclaves in which applications can execute securely and protect their data from unauthorised access or tampering. Confidential computing also provides remote and local attestations to provide evidence of the loaded logic and the state of the platform. 

TEEs are not all made equal and may exhibit different features. \cite{menetrey_attestation_2022} propose a list of state-of-the-art TEE features; crucial to implementing Honest computing are: Integrity, Freshness, Encryption and Remote Attestation. Naturally, while we assume an ideal TEE it is necessary over time to consider the security failures or existing architectures, as is the case for SGX for example (\cite{chen_sgxpectre_2019}), and to which extent these failures impact those crucial features. A sound Honest Computing implementation understands these risks and provides additional overlay mechanisms to safeguard processes and data.

For instance, to provide a sufficient level of robustness and redundancy, we envision an Honest Computing system to be comprised of a cluster of these TEEs responsible for executing application runtime within secure enclaves and a low-level software layer able to negotiate and share shards of private keys in memory to act as a buffer should the underlying hardware layer be compromised. Hardware sealing keys are platform-bound and different on every chip and, as such, guarantee that should the data be extracted out of a physical machine A, it would never be decryptable on physical machine B. This requires data synchronisation and transfer to be operated online between the nodes of a cluster. To realise this, Honest Computing implements multi-level consensus for its low-level primitives as well as supported runtime. In an ideal solution, we leverage TEEs from multiple manufacturers to support a broad scope of capabilities, limit the impact of any security problem affecting a given vendor and provide flexibility to the users. We can also leverage the cryptographic abilities of tamperproof HSMs in tandem with those of the TEEs (\cite{dib_hsmbased_2023}).

\subsection{Multi-level consensus}

Honest Computing relies on Trustless systems characteristics, pivotal for ensuring integrity, security, and transparency through:

\begin{itemize}
    \item \textbf{Decentralization:} Transparency is fostered through decentralization, reducing the risk of a single point of failure and nurturing trust among participants.

    \item \textbf{Consensus mechanisms:} Fault tolerance is achieved via consensus mechanisms, ensuring a unified truth even amidst individual agent failures (\cite{garg_generalized_2019}).

    \item \textbf{Distributed Ledger:} Distributed Ledger databases maintain an immutable historical record, preserving data integrity and a shared lineage among all involved entities.

\end{itemize}

In our Honest Computing approach, we distinguish between the data and process layers. The data layer provides the foundation for Honest Computing. It is organised as a set of TEE nodes sharing a ledger, featuring integrity validation throughout. We achieve this via the use of a distributed ledger, which requires an appropriate consensus mechanism. The process layer is concerned with the workflow implementation made by a stakeholder; to achieve honesty and fairness, an agreement between parties and users of the system will have to be reached. A minimum built-in consensus mechanism must be available to this process layer for Honest Computing to be effective.

Consensus mechanisms play a crucial role in ensuring the integrity and security of distributed ledger technologies (DLTs). They ensure that all participants in a decentralised network agree on the ledger's state, even in the absence of a trusted central authority. However, consensus mechanisms are also important in decision-making processes more broadly. We define them as being built into an Honest Computing systems software platform.

One attractive solution to the problem of process-level consensus is Shamir's secret sharing, also known as Shamir's key sharding. Shamir's secret sharing is a cryptographic technique that involves splitting a secret into multiple parts and distributing them among different participants in a network. To reconstruct the original secret, a minimum number of participants must combine their parts. This approach helps ensure that no single participant has complete control over the secret and reduces the risk of compromised cryptographic keys or passwords (\cite{shamir_how_1979}). Furthermore, it reduces the risk of compromised cryptographic keypairs, as an attacker would need to obtain multiple pieces from multiple participants to reconstruct the key. Third, it can help to ensure that the network remains operational even if some participants drop out or become unresponsive. As an extra consideration, we expect that a Shamir key sharding mechanism fit for Honest Computing would be constructed on top of key rotation mechanics to provide extra layers of safety. Shamir's secret sharing is used in various applications, such as securing cryptographic keys, protecting sensitive data, and ensuring distributed consensus in blockchain networks. In recent years, it has also been supplemented with integrity and validation (\cite{benzekki_verifiable_2017}).

\subsection{Distributed Ledger Technologies}

Distributed Ledger Technologies (DLTs) represent a class of digital systems designed to facilitate, record, and validate transactions across multiple locations or entities without the need for central oversight. These technologies offer decentralised, transparent, and secure methods of recording and managing data across a network of nodes. The most prominent form of DLT is blockchain. However, other variations and types of DLTs exist, each with its own unique features and applications. DLTs are prescribed as the storage heart of an Honest Computing system and ought to be used to preserve data and process code.

At the core of DLTs is the distributed ledger, a database that maintains a continuously growing list of records (blocks) linked together and secured through cryptography, where each node retains an identical copy of the ledger. To store, manage and retrieve this data efficiently in a cryptographically secure manner, ledgers can leverage specific data structures such as Merkle Patricia Tries (MPT). These data structures are usually structured as a tree, specifically a modified radix tree or trie, where each node represents a partial hash of its child nodes. The name ``Patricia'' stands for ``Practical Algorithm to Retrieve Information Coded in Alphanumeric'' and this data structure is space-efficient and allows for quick data integrity verification by storing hashes of node values. While Merkle Patricia Tries offer many advantages in terms of efficiency and security, they also pose challenges related to initial construction costs and complexities in handling large-scale data updates. Ongoing research aims to optimise these data structures further, making them more scalable and adaptable for various blockchain use cases beyond Ethereum (\cite{mardiansyah_multi-state_2023}).

DLTs traditionally use game theory to design mechanisms that incentivise participants to act in the best interest of the network as a whole. Game theory can help address some key challenges, such as the "tragedy of the commons" problem. By designing mechanisms (e.g., proof-of-work) that align participants' incentives with the network's goals, game theory can help DLTs achieve greater decentralisation, security, and efficiency. In another branch, in a proof-of-stake (PoS) blockchain network, game theory can be used to design a consensus mechanism that incentivises participants to validate transactions honestly and to penalise them for behaving maliciously. Our goal of providing a high throughput and ethical Honest Computing framework demands that we forgo these game theory-based mechanisms for the "proof-of-authenticity" or "proof-of-processor" that a TEE-based architecture offers.
By leveraging TEE local attestation capabilities, widely adopted consensus mechanisms, such as Raft, can be made Byzantine Fault Tolerant (BFT). It is crucial in the conception of DLT to ensure the utilisation of a BFT mechanism to safeguard systems from Byzantine failures.

Raft's ability to maintain consensus, elect leaders, and ensure log replication among nodes is advantageous in DLT environments where multiple nodes collaborate to maintain a shared and tamper-resistant ledger. While Raft is not commonly associated with blockchain networks that use Proof-of-Work or Proof-of-Stake mechanisms, its application can be found in permissioned or private blockchains, distributed databases, or other forms of DLTs where maintaining a decentralised yet more controlled environment is desirable.

\section{Discussion}\label{sec:discussion}

\subsection{Intrinsict benefits: An answer to data access revocation}

Data access revocation in computing refers to the complex challenge of effectively and promptly revoking access to sensitive or personal data that has been shared or provided to entities or systems, particularly in distributed and interconnected environments. The difficulty lies in ensuring that once access permissions are revoked, the data is no longer accessible or usable by unauthorised parties. This challenge is multifaceted and arises due to various reasons (\cite{politou_forgetting_2018}):

\begin{itemize}
    \item \textbf{Distributed Nature of Data:} Data is often stored and replicated across multiple systems, servers, or devices. Revoking access to data becomes challenging when it exists in diverse locations or has been shared across numerous platforms (\cite{tiwari_seccloudsharing_2018}).

    \item \textbf{Delayed Revocation:} Even if access rights are revoked centrally, propagating these changes across all systems and devices where the data resides might be time-consuming. During this propagation period, the data might still be accessible.

    \item \textbf{Data Copies and Backups:} Data might exist in backups or copies that are not immediately accessible or within the control of the data owner or administrator. Revoking access from all instances, including backups, adds complexity to the revocation process.

    \item \textbf{Compliance and Legal Constraints:} Regulatory compliance or legal obligations might necessitate retaining specific data even after access revocation, leading to complexities in ensuring complete data removal (\cite{vargas_mitigating_2018}).
\end{itemize}

Addressing the challenge of data access revocation requires comprehensive strategies and technological solutions. Implementing robust access control mechanisms, encryption, and identity management systems helps manage and revoke access more effectively. Furthermore, establishing protocols for data deletion and ensuring comprehensive audits and access rights monitoring can enhance the revocation process. However, overcoming these challenges in a fast-evolving digital landscape remains an ongoing endeavour, requiring continual innovation and policy considerations to mitigate risks associated with lingering data access after revocation.

\subsection{Remaining challenges and limitations}

\subsubsection{Right to be forgotten vs DLT}

Challenges arise when choosing DLT as a core technology to provide tamperproofness in that data is committed to a ledger that must persist forever to be effective. This means, in effect, that even while accounted for by the verifiable TEE system and even appropriately encrypted, the data remains available forever in the ledger. Depending on the perimeter and allowance of a given use case, stakeholders could address this problem by storing the data in external systems and only relying on the Honest Computing environment to maintain the key and access control functionality. It is worth noting that this could compromise deterministic behaviour relying on that data.

\subsubsection{Large data volume}
While permitting unprecedented speed of confidential computation, modern TEEs often have resource limitations such as restricted memory, computational power, or storage. This can impact the types and complexity of applications that can be securely run within the enclave, including in distributed systems. While our reference design for Honest Computing relies on Intel SGX TEE, other technologies, such as Intel Trusted Domain eXtensions (TDX), offering lower safety guarantees, could be leveraged to delegate large data processing workloads to task-specific hardware (\cite{sardar_formal_2023}). This is the choice made by NVIDIA with their confidential computing enabled H100 GPUs (\cite{nertney_confidential_2023}), helping to raise the challenge of large-scale AI computation. In these cases, more research needs to be done on the relationship between drivers and hosts and the ability to port drivers into a TEE directly.

\subsubsection{Complexity of implementation}
Developing secure applications for TEEs can be complex. Ensuring that applications properly utilise and interact with the TEE's security features without introducing vulnerabilities requires expertise and careful consideration. Data models utilised in confidential applications are still in their infancy and, to the best of our knowledge, do not feature native complex relational database systems but are limited to key-value store design. Intel TDX's ability to operate lift-and-shift could bring more existing software technologies to confidential computing environments, in some instances severely limiting the guarantees an integrated system offers due to the lack of deterministic execution guarantees.

\subsubsection{Use-case-dependent data loss threshold at the point of collection}

It remains essential to consider that areas of the pipeline described within the scope of Honest Computing exhibit limitations and provide appropriate assumptions. For example, under normal circumstances, we assume that there will inevitably exist scenarios where data gets lost due to transport failure followed by the collection agent's failure. There exist approaches based on meshed edge agents to deal with sensor networks that may never reach connectivity end-to-end (\cite{jenkins_sensor_2007}). Indeed, there are times when the collection is not possible at all (\cite{hayashi_data_2021}).

Also, Honest Computing does not cover the origination of sensor-based data itself due to the inability to estimate external factors. Some use cases may benefit from sensor trustworthiness evaluation research (\cite{lim_provenance-based_2010}), and other use cases are covered in ongoing research on metamaterials potentially able to generate cryptographic signatures while sensing (\cite{zhang_meta-optics_2023}).

\subsubsection{Lack of Open Hardware initiatives in the TEE space}

A notable challenge in trust execution environments (TEEs), particularly concerning hardware-based security solutions, is the absence of standardised open hardware specifications. 
However, the lack of open hardware standards in this domain poses several significant issues:

\begin{itemize}
    \item \textbf{Closed Architecture and Vendor Lock-in:} Many existing TEEs come from specific vendors and operate within closed architectures. This proprietary nature restricts interoperability and limits developers' options, potentially leading to vendor lock-in scenarios.

    \item \textbf{Transparency and Security Assurance:} Without open hardware standards, there is limited transparency into the hardware design and security mechanisms employed within TEEs. This lack of transparency hampers the ability of security researchers and experts to thoroughly review and audit the hardware for vulnerabilities or backdoors, raising concerns about trust and security assurance.

    \item \textbf{Innovation and Collaborative Development:} Open hardware standards encourage collaboration and innovation within the developer community. The absence of open standards stifles such collaborative efforts, hindering innovation in the TEE space.

    \item \textbf{Interoperability and Compatibility:} Open hardware standards promote interoperability among different hardware systems, thus fostering a more diverse ecosystem of secure hardware options.

    \item \textbf{Trust and Adoption:} Without standardised specifications and transparent designs, users may hesitate to fully trust the security claims of proprietary TEEs. It also creates circumstances where the root of trust becomes a point of failure contingent on the inability of a vendor to succumb to pressures.
\end{itemize}

Efforts are underway by various industry groups, open-source communities, and standards organisations to address this gap by advocating for more open and standardised hardware specifications in the TEE domain (\cite{lee_keystone_2020}).

\subsubsection{Cross-confidential computing environment communication}

Confidential computing presents a paradigm shift in ensuring data privacy and security by enabling data processing while it remains encrypted, even while in use. However, achieving cross-system communication in a confidential computing environment poses significant challenges due to the sensitive nature of encrypted data. Here are some of the key challenges:

\begin{itemize}
    \item \textbf{Interoperability:} Confidential computing often involves heterogeneous systems using diverse encryption schemes and protocols. Ensuring seamless communication while maintaining confidentiality requires standardised protocols and encryption methods, which can be challenging.

    \item \textbf{Key Management:}Coordinating key management in a way that allows cross-system communication without compromising confidentiality is a complex task.

    \item \textbf{Performance Overhead:} Encrypted cross-system communication and data processing can result in performance overheads due to increased complexity and additional encryption/decryption operations. 

    \item \textbf{Data Integrity and Authenticity:} Ensuring the integrity and authenticity of data transmitted between systems without compromising confidentiality is vital but challenging and requires additional cryptographic mechanisms.

    \item \textbf{Trust Establishment:} Ensuring that each system involved in the communication is trustworthy and adheres to security protocols presents a challenge, especially in distributed or decentralised systems.

    \item \textbf{Regulatory Compliance:} Maintaining compliance while enabling secure communication across systems adds complexity, especially when data crosses legal jurisdictions.

    \item \textbf{Lifecycle Management:} Handling data expiration, revocation of access, and proper disposal while ensuring confidentiality poses significant challenges in cross-system communication.
\end{itemize}

Addressing these challenges in confidential computing requires innovative solutions, including standardised encryption formats, robust key management practices, efficient cryptographic protocols, and secure communication channels. Moreover, collaborations between industry stakeholders and standardisation bodies are essential to develop interoperable frameworks and best practices for cross-system communication in a confidential computing environment while maintaining data confidentiality and integrity. While feature parity is not necessarily always available, research into the possibility of creating cross-technology protocols is underway (\cite{antonino_flexible_2023}).

\subsection{Additional considerations}

\subsubsection{Homomorphic encryption}

Homomorphic encryption is a cryptographic technique that allows computations to be performed on encrypted data with the results remaining in encrypted form. It represents a significant advancement in preserving data privacy and security, especially when sensitive information needs to be processed or analysed without exposing the raw data.

However, homomorphic encryption faces challenges such as performance overhead due to computational complexity, slower processing speed compared to plaintext operations, and limitations on the types and complexity of computations that can be performed efficiently (\cite{brenna_tfhe-rs_2022}).

Our reference specification acknowledges these potentials but discounts the use of homomorphic encryption systems at this point due to their generally poor performance and implementation complexity. It is also worth noting that the computational integrity of homomorphic encryption algorithms is still lacking (\cite{viand_verifiable_2023}).

\subsubsection{Post-Quantum resitance}
Post-quantum resistance is becoming increasingly important in cryptography due to the potential threat posed by quantum computers to traditional cryptographic algorithms.

In 2022, The National Institute of Standards and Technology (NIST) announced its selection of four finalist algorithms for post-quantum cryptography, including Falcon, Saber, Dilithium, and Rainbow (\cite{nist_nist_2022}). They are still undergoing extensive analysis and evaluation before they can be approved for use. As such, staying up-to-date on the latest developments in post-quantum cryptography is crucial in ensuring that any cryptographic systems are updated to use NIST-approved post-quantum cryptographic algorithms once they become available. Current widely spread symmetric systems such as AES-256 are still safe against known quantum attacks and algorithms (\cite{sharma_study_2023}).

\subsubsection{Poor Software practices}
While an Honest Computing system can achieve a very high level of privacy and confidentiality, it cannot be a substitute for correct software review practices. Recent studies in the US have shown that poor software development practices are costly and impactful when data safety is compromised (\cite{krasner_cost_2020}). Software designed with features permitting data extraction must be carefully considered, and regulatory guidelines may need to be put in place to ensure proper communication around software behaviour, management, and updates. It is worth noting that should software running in Honest Computing environments be compromised, the integrity, tamper-evidentness and attestability features of Honest Computing systems would, at the very least, allow the detection of these failures. More recent research proposes a new way for policymakers to evaluate software security quality when built on top of an Honest Computing system (\cite{larios-vargas_dasp_2023}).

\subsubsection{Requirements for Differential Privacy}

Differential privacy is a set of techniques in the field of data privacy that aims to enable the analysis of datasets while protecting the privacy of individual data points (\cite{aziz_exploring_2023}). The central idea behind differential privacy is to add a controlled amount of noise to the data or query results in a way that ensures that the contribution of any single individual's data is indistinguishable from the noise. Differential privacy is particularly relevant in data analysis, machine learning, and statistical databases, where the goal is to derive meaningful insights from sensitive datasets without compromising the privacy of individuals contributing to those datasets. While it is effective on a case-by-case basis, it has some weaknesses. One weakness is that it can reduce the accuracy of the data, making it difficult to draw meaningful conclusions (\cite{jain_price_2023}). Honest Computing can support Differential privacy implementations. However, it is left at the discretion of stakeholders if and when it is relevant.

\subsection{Domains of application}
Honest Computing is specified as a low-level technology on top of which applications get built. In consideration of what it enables, it is immediately suited to use cases in a variety of fields, examples of which are: Healthcare (\cite{shae_design_2017}), Finance and Banking (\cite{daligny_cental_2022}), Trading (\cite{xue_blockchain-based_2023}), Supply Chain Management(\cite{herbe_how_2023}), Notarisation or Regulation (AML/CFT~\cite{pocher_privacy_2022}).

\section{Conclusion}\label{sec:conclusion}

In this paper, we introduced the concept of Honest Computing, rooted in trustless computing and confidential computing paradigms, as a promising solution to address the intricate challenges within existing regulatory frameworks, particularly in the realm of sensitive data processing, lineage and provenance in automated decision-making systems. The limitations of current regulations, often stemming from a lack of technical readiness and detailed protocols, hinder their ability to provide comprehensive guidelines for compliance, transparency, and accountability.

Honest Computing offers a technological foundation that has the potential to bridge regulatory gaps by enabling the creation of auditable, validatable, and enforceable rules. By leveraging trustless computing principles and confidential computing techniques, it promises to enhance transparency and effectiveness and provides a scalable solution for policymakers. Honest Computing holds the potential to usher in a new era of regulatory frameworks by empowering regulators with the tools to establish guidelines that are not only principle-based but underpinned by robust technical protocols. In this way, policymakers stand to benefit significantly, gaining the means to navigate the complexities of data life cycles while ensuring compliance, fairness, and ethical conduct across various sectors.




\begin{Backmatter}

\paragraph{Acknowledgments}
We would like to thank our reviewers for their time and feedback. We would also like to extend our thanks to the research teams at the Data Science Institute at Imperial College London. We are also grateful for the industry insights provided by the people at Secretarium Ltd, particularly Bertrand Foing and Cedric Wahl. The authors used Microsoft 365 Copilot 25997.1010 and Grammarly to assist with drafting and proofing this article via their Microsoft Word and Chromium integration.

\paragraph{Funding Statement}
This work was partially supported by Innovate UK (Project Reference 72834) and the Data Science Institute at Imperial College London.

\paragraph{Competing Interests}
The authors declare no competing interests exist.

\paragraph{Data Availability Statement}
All the code relevant to the exploration in this report is available through GitHub repositories at \verb+https://github.com/dsi-icl+, \verb+https://github.com/secretarium+ and \verb+https://github.com/klave-network+.

\paragraph{Author Contributions}
 All authors approved the final submitted draft. Conceptualization: F.G., A.O.; Project administration: F.G; Methodology: F.G., A.O., E.B.; Writing original draft: F.G., A.O.; Writing—review and editing: F.G., A.O., E.B.; Supervision: Y.G.




\printbibliography






\end{Backmatter}

\end{document}